\documentstyle[12pt,epsf,epsfig,amsmath,amsfonts,color]{article}

\def\be{\begin{equation}}
\def\ee{\end{equation}}
\def\ba{\begin{eqnarray}}
\def\ea{\end{eqnarray}}

\def\bdm{\begin{displaymath}}
\def\edm{\end{displaymath}}
\def\la{~\mbox{\raisebox{-.6ex}{$\stackrel{<}{\sim}$}}~}
\def\ga{~\mbox{\raisebox{-.6ex}{$\stackrel{>}{\sim}$}}~}
\def\bq{\begin{quote}}
\def\eq{\end{quote}}

 at 10truept

\newcommand{\beq}{\begin{equation}}
\newcommand{\eeq}{\end{equation}}
\newcommand{\bea}{\begin{eqnarray}}
\newcommand{\eea}{\end{eqnarray}}
\newcommand{\beqa}{\begin{eqnarray}}
\newcommand{\eeqa}{\end{eqnarray}}

\def\la{~\mbox{\raisebox{-.6ex}{$\stackrel{<}{\sim}$}}~}
\def\ga{~\mbox{\raisebox{-.6ex}{$\stackrel{>}{\sim}$}}~}

\def\ltap{\ \raise.3ex\hbox{$<$\kern-.75em\lower1ex\hbox{$\sim$}}\ }
\def\gtap{\ \raise.3ex\hbox{$>$\kern-.75em\lower1ex\hbox{$\sim$}}\ }
\def\gl{\ \raise.5ex\hbox{$>$}\kern-.8em\lower.5ex\hbox{$<$}\ }
\def\roughly#1{\raise.3ex\hbox{$#1$\kern-.75em\lower1ex\hbox{$\sim$}}}

\begin{document}

\thispagestyle{empty}
\begin{flushright}
September 2014
\end{flushright}
\vspace*{1cm}
\begin{center}
{\Large \bf `The End'}

\vspace*{1.1cm} {\large Nemanja Kaloper$^{a, }$\footnote{\tt
kaloper@physics.ucdavis.edu} and Antonio Padilla$^{b, }$\footnote{\tt
antonio.padilla@nottingham.ac.uk} }\\
\vspace{.3cm} {\em $^a$Department of Physics, University of
California, Davis, CA 95616, USA}\\
\vspace{.3cm} {\em $^b$School of Physics and Astronomy, 
University of Nottingham, Nottingham NG7 2RD, UK}\\

\vspace{1.5cm} ABSTRACT
\end{center}
Recently we proposed a mechanism for sequestering the Standard Model vacuum energy that predicts that the universe 
will collapse. Here we present a simple mechanism for bringing about this collapse, employing a scalar field whose potential is linear and becomes negative, providing the negative energy density required to end the expansion. The slope of the potential is chosen to allow for the expansion to last 
until the current Hubble time, about $10^{10}$ years, to accommodate our universe. Crucially, this choice  is technically natural due to a shift symmetry. Moreover, vacuum energy sequestering selects {\it radiatively stable} initial conditions for the collapse, which guarantee
that  immediately before the turnaround the universe is dominated by the linear potential which drives an epoch of accelerated expansion for at least an efold. Thus a single, technically natural choice for the slope ensures that the collapse is imminent and  is preceded by the current stage of cosmic acceleration, giving a new answer to the `Why Now?' problem. 

\vfill \setcounter{page}{0} \setcounter{footnote}{0}
\newpage

In \cite{kp1,kp2} we proposed a mechanism for sequestering the matter sector vacuum energy from gravity.
The idea is that all the matter sector scales are functions of the $4$-volume element of the 
universe $\int d^4 x \sqrt{g}$. 
Picking the specific functional dependence on $\int d^4 x \sqrt{g}$, reflecting the scaling of dimensional quantities, 
ensures that the vacuum energy contributions are invisible to gravity at any and all orders of the loop expansion in the matter sector.
While this `cancellation' of the vacuum energy does not directly protect from graviton loops, it is radiatively stable to matter corrections, 
evading the Weinberg no-go theorem \cite{wein} (see also \cite{zeldovich}). As a bonus, the mechanism renders the contributions from phase
transitions automatically small at late times. Since our mechanism is locally Poincar\'e invariant and diffeomorphism invariant, it does not introduce any new local degrees of freedom, representing a minimal modification of General Relativity by adding two global constraints. 

For the matter sector scales to be nonzero, the universe 
should be finite in spacetime, collapsing in the future. 
This requires dynamics which can turn expansion into contraction in an FRW geometry. A simple possibility is to have a field with a potential that is not bounded from below, or at least is `sufficiently negative' (see, e.g. \cite{wilczek,lindelinder} for earlier considerations). If the 
negative potentials compensate the positive energy density of the rest, the collapse will occur, and the field theory will not be scale-invariant but will have a nonzero mass gap.  Having a negative potential that can destabilize the universe seems to be physically feasible. After all, the Standard Model Higgs appears to be an example of a field with such a potential \cite{giudice}, albeit with a very delayed instability \cite{sala}. 

After vacuum energy cancellation there remains a residual, radiatively stable cosmological constant whose value must be smaller than the value of the turnaround potential if collapse is to occur.  By taking the turnaround potential to be a linear function, it too is radiatively stable.
This is because 
a linear potential is exactly shift symmetric in our framework, sufficing not only to protect its form but also the numerical value of its slope. Once set, to the leading order this potential remains  the same to all orders in the loop expansion in the matter sector. This is crucial since it allows us to make a {\it technically natural} choice of the slope, $m_{slope}^3 \simeq M_{Pl} H_0^2 \simeq 10^{-39} (\text{eV})^3$,  that ensures that collapse will occur after a present Hubble epoch. Further, the vacuum energy sequestering dynamically selects radiatively stable initial conditions that {\it guarantee} this turnaround will be preceded by a period of accelerated expansion at the observed scale of dark energy, lasting for at least an efold. 
This is  a new solution to the `Why Now?' problem\footnote{Although the linear potential has been discussed before in GR to address the coincidence problem \cite{Avelinonow}, there the radiative corrections to $\Lambda$ require fine tunings, unlike in our proposal, as we will discuss below.}
 \cite{whynow}.
It also confirms that $w_{DE} \simeq -1$ is a transient, as we claimed in \cite{kp1,kp2}, and may be 
observable, along the lines of \cite{lindelinder}. 

As defined in \cite{kp1,kp2}, we start with the theory given by
\be
S= \int d^4 x \sqrt{g} \left[ \frac{M^2_{Pl}}{2} R  - \Lambda - {\lambda^4} {\cal L}(\lambda^{-2} g^{\mu\nu} , \Phi) \right] +\sigma\left(\frac{ \Lambda}{\lambda^4 \mu^4}\right) 
 \, . ~~
\label{action}
\ee 
where all ``protected'' matter couples minimally to the rescaled metric $\tilde g_{\mu\nu}=\lambda^2 g_{\mu\nu}$, and as before $\lambda$ sets the hierarchy between the matter scales and the Planck scale, since $m_{phys}/M_{Pl} \propto \lambda m/M_{Pl}$,
 where $m_{phys}$ is the physical mass scale and $m$ is the bare mass in the Lagrangian. The parameters $\Lambda$ and $\lambda$ are global Lagrange multipliers, constants in space and time that should be varied over to minimise the action, yielding global constraints
 \be
  \frac{\sigma'}{\lambda^4\mu^4} = \int d^4x \sqrt{g} \, , ~~~~~~~~~~~~~~~~ 4\Lambda \frac{ \sigma' }{\lambda^4\mu^4} 
= \int d^4x \sqrt{g} \,  T^{\mu}{}_\mu \, ,
\label{varsl} 
\ee
where $ T_{\mu\nu}=\frac{2 \lambda^4}{\sqrt{\tilde g}} \frac{\delta }{\delta \tilde g^{\mu\nu}}\int d^4 x \sqrt{\tilde g} {\cal L}(\tilde g^{\mu\nu} , \Phi)$ is the conserved stress-energy in the `physical' frame, where the matter sector is canonically normalized, and $\sigma' = \frac{d\sigma(z)}{dz} \ne 0$
Thus the function $\sigma$ fixes the matter scales as functions of $\int d^4x \sqrt{g}$. 
Phenomenological arguments favor odd functions which are exponential for large arguments in order to desensitize low energy 
particle physics from dependence on cosmological initial conditions, while the scale $\mu$ is a fixed external scale 
determined by same \cite{kp1,kp2}.  The full protected matter Lagrangian 
\be \label{L}
 {\lambda^4} {\cal L}(\lambda^{-2} g^{\mu\nu} , \Phi) = {\lambda^4} \hat {\cal L}(\lambda^{-2} g^{\mu\nu} , \Phi) + \frac{\lambda^2}{2} (\partial \varphi)^2 
+ \lambda^4 V(\varphi) 
 \ee
includes the Standard Model  fields (along with possible extensions) in $\hat {\cal L}$, and a scalar  field $\varphi$ which is the collapse `trigger'  controlling the turnover of cosmic expansion. The vacuum energy cancellation\footnote{Even if we ignore graviton loops, the Weinberg no-go theorem \cite{wein} precludes vacuum energy adjustment in a standard EFT. It is important that the QFT regulator must also couple minimally to $\tilde g_{\mu\nu}$ to ensure the cancellation of $\lambda$ in loop logarithms \cite{kp1,kp2}.} to all orders in the protected  matter sector loop expansion follows from diffeomorphism invariance of the theory, since the full effective Lagrangian computed from 
$\sqrt{g} \lambda^4 {\cal L}(\lambda^{-2} g^{\mu\nu} , \Phi)=\sqrt{\tilde g} {\cal L}(\tilde g^{\mu\nu} , \Phi)$, 
still couples to $\tilde g_{\mu\nu}$ \cite{selft}. 

Integrating out the global variables $\Lambda$ and $\lambda$, the field equations become \cite{kp1,kp2} 
\be
M_{Pl}^2 G^\mu{}_\nu= T^\mu{}_\nu-\frac{1}{4} \delta^\mu{}_\nu \langle T^\alpha{}_\alpha \rangle \, .
\ee
where  the historic average is denoted $\langle Q\rangle ={\int d^4 x \sqrt{g} \, Q}/{\int d^4 x \sqrt{g}}$.  If we take the effective matter Lagrangian ${\cal L}_\textrm{eff}$ calculated to any order in loops, and split it into the renormalised vacuum energy (classical and quantum) $\tilde V_{vac}=\langle 0|{\cal L}_\textrm{eff}|0\rangle$, and local excitations $\Delta {\cal L}_\textrm{eff}$, we can write $ T^{\mu}_{\nu}= -V_{vac}\delta^{\mu}_{\nu}+\tau^\mu_\nu$, where $V_{vac}=\lambda^4 \tilde V_{vac}$ and $\tau_{\mu\nu}=\frac{2 }{\sqrt{ g}} \frac{\delta }{\delta g^{\mu\nu}}\int d^4 x \sqrt{ g} \lambda^4 \Delta {\cal L}_\textrm{eff}(\lambda^{-2}g^{\mu\nu} , \Phi)$. The vacuum energy completely drops out of the field equations, leaving us with 
\be \label{eeqs}
M_{Pl}^2 G^\mu{}_\nu= \tau^\mu{}_\nu-\frac{1}{4} \delta^\mu{}_\nu \langle \tau^\alpha{}_\alpha \rangle \, .
\ee
A residual cosmological constant $\Lambda_\textrm{eff}=\frac14 \langle  \tau^\alpha{}_\alpha \rangle$ remains, and corresponds to a radiatively stable renormalized
cosmological constant operator, after subtracting the divergent part. 
It must be {\it measured} as in the case of any divergent quantities
in quantum field theory. Our procedure amounts to enforcing the result of the measurement by using the whole universe as the
detector \cite{kp2}. It is protected by approximate scaling and shift symmetries of the theory 
that control the cancellation \cite{kp1,kp2}.

Because $\lambda$ controls the physical scales in ${\cal L}$, $m_{phys} = \lambda m$, it must be nonzero. By the first of 
Eqs. (\ref{varsl}), $\int d^4 x\sqrt{g}$ must also be finite \cite{kp1,kp2}. 
For this to happen the cosmic expansion must be halted, and contraction must begin. This is the job for the collapse trigger $\varphi$ in (\ref{L}). Its potential $V(\varphi)$ must be negative for at least some range of $\varphi$ to halt the expansion \cite{lindelinder,waldnohair}. It cannot be constant: if so it would have been cancelled along with the rest of the vacuum energy in (\ref{action}). An arbitrary potential would be subject to radiative corrections, which change both its form and the numerical values of its scales. Dealing with
this would restore some of the tunings of the vacuum energy sector which we are striving to evade. 

In the case of a {\it linear potential} $V({\varphi}) = m_*^3 \varphi + V_0$, however, there is a powerful protection mechanism from radiative corrections. 
It is the approximate scaling and shift symmetries of (\ref{action}) that  ensure the cancellations of the vacuum energy in the first 
place \cite{kp1,kp2}. Indeed, the shift $\varphi \rightarrow \varphi + {\cal C}$ changes the  full matter Lagrangian by ${\cal L} \rightarrow {\cal L} + m_*^3 {\cal C}$. This can be absorbed by the shift of the global variable $\Lambda \rightarrow \Lambda - \lambda^4 m_*^3 {\cal C}$ so that the bulk action in (\ref{action}) is invariant. The global term changes by 
$\sigma \rightarrow \sigma + \sum_{n=1}^\infty \frac{(-1)^n}{n!} {\sigma^{(n)}}|_{{\cal C}=0} \bigl(\frac{m_*^3 {\cal C}}{\mu^4}\bigr)^n$, but this has no effect on the scales in the bulk terms in perturbation theory such as 
$m_*$. The only effect arises from a variation of $\lambda$ induced by a change of the constraint equations (\ref{varsl}), renormalizing 
the physical scales by the $\lambda$ variation. Whenever $\frac{m_*^3 {\cal C}}{\mu^4} \ll 1$, these variations are small, as generically occurs for small scales $m_*$ and large cutoffs $\mu \sim M_{Pl}$. Hence, not only is the linear form of $V(\varphi)$ protected, but its physical slope $m_{slope}^3 = \lambda^3 m_*^3$ is perturbatively stable too. We can choose it to be whatever we like because its small values are technically natural. As we will see, a solution to the ``Why now?" problem follows from the choice $m_{slope}^3 \simeq M_{Pl} H_0^2$.

Let us now study the dynamics of the linear potential in the sequestering framework. Note that now the residual cosmological constant,  $\Lambda_\textrm{eff}=\frac14 \langle  \tau^\alpha{}_\alpha \rangle$,  is irrelevant by itself, since it can be `gauged away' by a shift\footnote{By a shift of $\varphi$, we can also remove $V_0$ from all local equations; it will appear in the constraints (\ref{varsl}), but for asymptotically exponential $\sigma$, its effects are negligible. } of $\varphi$. 
To see this explicitly we write   $\tau^\mu{}_\nu = \hat \tau^\mu{}_\nu + \partial^\mu \varphi \partial_\nu\varphi - \frac12 \delta^\mu{}_\nu (\partial \varphi)^2 - \delta^\mu{}_\nu m_{slope}^3 \varphi$, where $\hat \tau^\mu{}_\nu$ is the stress-energy of all protected matter sector fields other than $\varphi$, contained in $\hat {\cal L}$. It follows that $\langle \tau^\mu{}_\mu \rangle = \langle \hat \tau^\mu{}_\mu \rangle - \langle (\partial \varphi)^2\rangle - 4 m_{slope}^3 \langle \varphi \rangle$, and so the RHS of Eq. \ref{eeqs} is given by
\be
\tau^\mu{}_\nu - \frac14 \delta^\mu{}_\nu \langle \tau^\alpha{}_\alpha \rangle = \hat \tau^\mu{}_\nu - \frac14 \delta^\mu{}_\nu \langle \hat \tau^\alpha{}_\alpha \rangle
+ \partial^\mu \varphi \partial_\nu\varphi - \frac14 \delta^\mu{}_\nu \Bigl( 2(\partial \varphi)^2 - \langle (\partial \varphi)^2 \rangle \Bigr)
- \delta^\mu{}_\nu m_{slope}^3 (\varphi - \langle \varphi \rangle )  \, .
\ee
This stress-energy tensor is manifestly invariant under $\varphi \rightarrow \varphi + {\cal C}$. In contrast, the residual cosmological constant is not invariant. It transforms as
\be
\langle \tau^\mu{}_\mu \rangle \rightarrow \langle \tau^\mu{}_\mu \rangle -  4 m_{slope}^3 {\cal C} \, .
\ee
Shifting the origin of the $\varphi$ direction we can set $\langle \tau^\alpha{}_\alpha \rangle$ to any arbitrary value, and in particular, we can choose it to vanish\footnote{The scalar kinetic energy dominates near the big crunch \cite{Avelinoseq} and so $\langle \tau^\mu{}_\mu \rangle$ diverges logarithmically. So does the curvature. Hence one 
must terminate the geometry at Planckian densities, which renders the contributions from the singularity small in big universes \cite{kp2}.}.

We now turn to cosmological mechanics and  take
spatially 
closed 
FRW,
$ds^2 = -dt^2 + a^2(t) d \Omega_3$
as the background geometry, describing a compact
universe that underwent a stage of rapid inflation  \cite{kp1,kp2}. As noted above, we will take the $\varphi$ gauge such that the potential is $m_{slope}^3 \varphi$, with the initial value of $\varphi$ to 
be determined aposterior to satisfy the gauge constraint $\langle  \tau^\alpha{}_\alpha \rangle=0$ \cite{kp1,kp2}. 
With $H = \dot a/a$, the field equations (\ref{eeqs}) reduce to
%
\bea
&3M_{Pl}^2 \left(H^2 +\frac{1}{a^2}\right) = \rho + \frac{\dot \varphi^2}{2} + m_{slope}^3 \varphi , ~~~~~~ 3M_{Pl}^2 \left(\dot H-\frac{1}{a^2}\right) = - \frac32 (\rho + p) - \frac32 \dot \varphi^2,& \nonumber \\   ~~~~~~
&\ddot \varphi + 3H\dot \varphi + m_{slope}^3 = 0 .&
\label{eqsfrw} 
\eea
The energy density $\rho$ of the protected sector matter other than $\varphi$ obeys $\dot \rho + 3H (\rho + p) = 0$, where
$p \ne - \rho$: it is {\it not} vacuum energy, which completely cancels from the source.
The gauge constraint $\langle  \tau^\alpha{}_\alpha \rangle=0$ yields 
\be
  \langle \dot \varphi^2 \rangle -4 m_{slope}^3 \langle \varphi \rangle
+ \langle 3p - \rho\rangle=0 \, .
\label{constrcrucial}
\ee
and  will play a crucial role in what follows.
As stressed above, we have canonically normalized all modes, absorbing $\lambda$ into the 
dimensionful quantities in ${\cal L}$.

Starting with (\ref{eqsfrw}) we first prove that the linear potential forces an expanding universe to eventually collapse. We do so by contradiction: suppose collapse does {\it not} happen, then at late times the scale factor $a$ must be strictly positive. If it were also finite, we could have written it asymptotically as 
$a = a_{end} + f(1/t)$ for some suitable\footnote{The function should be continuous in a neighbourhood of the origin. It should also be twice differentiable there, except possibly at the origin itself.} function $f$ obeying $f(0)=0$. Then 
\be
H \rightarrow - \frac{g(1/t)}{t \, a_{end}} \, , ~~~~~~~~~~~~~~~~~~  \dot H \rightarrow \frac{g(1/t) + h(1/t)}{t^2 \, a_{end}} \, ,
\ee
with $g(x) = x f'(x)$ and $h(x) = xg'(x)$. So asymptotically $\dot H \rightarrow 0$, implying that 
$\dot \varphi \rightarrow$ constant, and so $\ddot \varphi \rightarrow 0$. It also follows that $H \rightarrow 0$, which is in contradiction with the last equation in  (\ref{eqsfrw}) when $m_{slope} \neq 0$. If, on the other hand, $a_{end}$ diverges, we again
 infer that $H \rightarrow 0$, and by the Friedmann equation in (\ref{eqsfrw}), also
$\frac12 \dot \varphi^2 + m_{slope}^3 \varphi \rightarrow 0$. From the vanishing of $\dot H$ we
again
 infer $\dot \varphi \rightarrow 0$ and so we must also have $\varphi \rightarrow 0$. We now arrive at our
 final 
  contradiction because the last equation of (\ref{eqsfrw}) implies $\ddot \varphi \rightarrow - m_{slope}^3 \ne 0$, which is impossible by analyticity when $\varphi, \dot \varphi \rightarrow 0$. Hence the universe {\it must} collapse.

Can the collapse driven by a linear potential be delayed for long enough to approximate our universe? 
This requires initial values of $\varphi_{in}$ for which $V(\varphi_{in}) > 0$, and a potential slope gradual enough such that the epoch
of $V(\phi) > 0$ is sufficiently long. Since the slope is technically natural, its chosen value will be stable. If we choose it to be $m_{slope}^3 \simeq M_{Pl} H_0^2$, the ensuing universe will expand until a time $\sim H_0^{-1}$, and undergo a short stage of accelerated expansion just before the collapse. In contrast to GR where other possibilities can be realized, the vacuum energy sequestering dynamically predicts this
outcome via the gauge constraint (\ref{constrcrucial}), which pick the special, radiatively stable initial conditions for the scalar field $\varphi$.

Now we prove this. For $m_{slope}^3$ small, unless the trigger field $\varphi$ is initially many orders of magnitude larger than $M_{Pl}$,
its energy density will be subleading to other matter sources. Ignoring its contributions to the gravitational equations in (\ref{eqsfrw}), in an expanding  power law FRW background ($a \sim t^p$ for $p>0$), at large wavelengths 
the field equations give
$\dot \varphi = \frac{\varphi_1}{t^{3p}} - \frac{m_{slope}^3 t}{3p+1}$ and $\varphi = \varphi_0 + \frac{\varphi_1}{(1-3p) t^{3p-1}} - \frac{m_{slope}^3 t^2}{2(3p+1)}$. The attractor values of $\varphi, \dot \varphi$ are  
\be
\varphi = \varphi_0 - \frac{m_{slope}^3 t^2}{2(3p+1)} = \varphi_0 - \frac{m_{slope}^3}{2(3p+1)} \left(\frac{p}{H}\right)^2 \, , ~~~~~~~~~
\dot \varphi =  - \frac{m_{slope}^3 t}{3p+1} =  - \frac{m_{slope}^3 }{3p+1} \frac{p}{H} \, .
\label{attractor}
\ee
So $\varphi$ is speeding up towards more negative values. It will inevitably begin to dominate the dynamics triggering collapse. Suppose 
that it begins to dominate while $\varphi > 0$ at a scale $H_{\dagger}$, where $\varphi_\dagger = \varphi_0 - \frac{m_{slope}^3}{2(3p+1)} \left(\frac{p}{H_\dagger}\right)^2$
and $\dot \varphi_\dagger = - \frac{m_{slope}^3}{3p+1} \frac{p}{H_\dagger}$. Until such time, its total variation is
\be
\frac{\varphi_\dagger-\varphi_{in}}{M_{Pl}} = \int^{t_\dagger}_{t_{in}} \frac{dt}{M_{Pl}}  \, \dot \varphi 
\simeq - {\cal O}(1) 
\frac{m^3_{slope} M_{Pl}}{M_{Pl}^2 H_\dagger^2} \, .
\label{deltaphi}
\ee
Since we chose $m_{slope}^3 \simeq M_{Pl} H_0^2$, where $H_0 \la H_\dagger$, the field displacement is sub-Planckian during the whole past history of the universe up until $\varphi$ domination,  $\varphi_{in}-\varphi_\dagger< M_{Pl}$.  

Further, at the onset of $\varphi$ domination, 
$3 M_{Pl}^2 H_\dagger^2 \simeq \rho_\dagger(\varphi)\equiv \frac{\dot \varphi_\dagger^2}{2} + m_{slope}^3 \varphi_\dagger $. Our choice of small $m_{slope}^3 \simeq M_{Pl} H_0^2$ sets the ratio of kinetic to total energy of $\varphi$ to be 
$ \frac{\frac12 \dot \varphi_\dagger^2}{\rho_\dagger(\varphi)} = {\cal O}(1) \frac{m_{slope}^6}{M_{Pl}^2 H_\dagger^4} < 1$. So 
the energy density is dominated by its
potential energy, and $m_{slope}^3  \varphi_\dagger \simeq 3 M_{Pl}^2 H_\dagger^2$, or equivalently 
$\varphi_\dagger \simeq \frac{M_{Pl} H_\dagger^2}{m_{slope}^3} M_{Pl} > M_{Pl}$. Similarly, the ratio of kinetic and potential energy in $\varphi$ is $\frac{\dot \varphi^2_\dagger/2}{m_{slope}^3 \varphi_{\dagger}} \simeq {\cal O}(1) \frac{m_{slope}^6}{M_{Pl}^2 H_\dagger^4} < 1$, and 
since $\varphi_\dagger > M_{Pl}$, we conclude that the trigger field is automatically in slow roll once it begins to dominate!  Indeed, the slow roll parameters for the linear potential are 
\be
\epsilon = \frac{\dot \varphi^2_\dagger}{2 m_{slope}^3 \varphi_{\dagger}} < 1 \, , ~~~~~~~~~~~~~~~
\eta = \frac{\ddot \varphi_\dagger}{H_\dagger \dot \varphi_\dagger} \simeq \frac12 \frac{M_{Pl}^2}{\varphi_\dagger^2} < 1 \, ,
\label{slowroll}
\ee
by our choice of $m_{slope}$. 

The evolution outlined above is generic provided $\varphi_{in} > M_{Pl}$ thanks to the sub-Planckian field displacements leading up to $\varphi$ domination. The gradual change of the linear potential guarantees that the collapse will be delayed to very late times, $t > H_0^{-1}$, by the fact that the field $\varphi$ is in slow roll. What is more, before the collapse, the universe undergoes a stage of accelerated expansion. This essentially  lasts until the turnover, which occurs when $m_{slope}^3 \varphi \simeq - \frac12 \dot \varphi^2$. We can estimate the scale at which this happens by using the slow roll description of $\varphi$ evolution, 
\be
3M_{Pl}^2 H^2 =m_{slope}^3 \varphi \, , ~~~~~~~~~~~~~~~ -3H\dot \varphi +m_{slope}^3=0 \, ,
\ee
yielding $\frac23 \varphi^{3/2} = - \sqrt{\frac{m^3_{slope}}{3}} M_{Pl} (t - t_\dagger) + \frac23 \varphi_\dagger^{3/2}$. Slow roll approximation breaks down when $\varphi  \la M_{Pl}$, i.e. when $t-t_\dagger \simeq  {\cal O}(1)  \sqrt{M_{Pl}/m_{slope}^3}$. After that, the field quickly runs to negative values to bring about the turnover. To a good approximation, the turnover occurs at a time $t_{turnover} \sim 1/H_{age}$ where 
\be
\frac{1}{H_{age}}  \simeq \frac{1}{H_\dagger} + {\cal O}(1) \sqrt{\frac{M_{Pl}}{m_{slope}^3}} \ga \frac{1}{H_0} \, ,
\label{age}
\ee
which gives  the number of efolds of acceleration preceding the collapse to be ${\cal N} \sim {H_\dagger}/{H_0}$. 
The total variation of $\varphi$ throughout the expanding phase can be estimated an extension of Eq. (\ref{deltaphi}) 
which gives
\be
\frac{\varphi_{turnover}-\varphi_{in}}{M_{Pl}} \simeq - {\cal O}(1) 
\frac{m^3_{slope} M_{Pl}}{M_{Pl}^2 H_{age}^2} \simeq - {\cal O}(1) \frac{H_0^2}{H_{age}^2}  \, .
\label{deltaphia}
\ee
Clearly  $\varphi_{in}-\varphi_{turnover} \ga M_{Pl}$, and so most of that variation accumulated near the turnover. 

In standard GR, the evolution we have just described arises from a small subset of possible initial conditions. There are many more reasonable initial conditions and they lead to very different cosmic eschatology. 
However, our framework is far more restrictive than standard GR. In our case the only permissible initial conditions are those that {\it guarantee} the evolution along the lines described above. Indeed,  recall the constraint (\ref{constrcrucial}) that restricts the initial conditions. Since the universe is spatially compact, and collapses, the right-hand side is dominated by the contributions from near the turning point. By arguments similar to those explained in \cite{kp2}, it follows that $\langle \dot \varphi^2 \rangle -4 m_{slope}^3 \langle \varphi \rangle \simeq \rho_{turnover}(\phi) \simeq -{\cal O}(1) m^3_{slope} \varphi_{turnover}$ and 
$ \langle 3p - \rho\rangle \simeq - {\cal O}(1) M_{Pl}^2 H^2_{age}$, so the constraint  (\ref{constrcrucial}) gives 
\be
\frac{\varphi_{turnover}}{M_{Pl}} \simeq  - {\cal O}(1) \frac{M_{Pl} H_{age}^2}{m_{slope}^3} \simeq -{\cal O}(1) \frac{H_{age}^2}{H_0^2} \ga -1\, .
\label{constcond}
\ee
This and Eq. (\ref{deltaphi}) force the initial value of $\varphi$ to be trans-Planckian: 
\be
\varphi_{in} \simeq \left( {\cal O}(1)    \frac{H_0^2}{H_{age}^2}+ {\cal O}(1) \frac{H_{age}^2}{H_0^2}\right) M_{Pl} \ga M_{Pl} \, .
\ee
This is precisely the initial condition that leads to cosmological collapse in the imminent future, preceded by a period of slow roll and accelerated expansion. 
Other initial values of $\varphi_{in}$  that could have led to different cosmic evolutions are excluded in our setup. They are {\it dynamically impossible}. Thus in our framework a collapsing universe, with the collapse triggered by a field with a linear potential, {\it must} undergo a stage of accelerated expansion prior to collapse, and after the onset of scalar field domination. 

Let us summarize. Here we have presented a mechanism to trigger cosmological collapse in the framework for vacuum energy sequestering  recently proposed  in \cite{kp1,kp2}. It is  a field theory with a linear potential, whose form and slope are protected by a shift symmetry, and so are
technically natural. To delay the collapse the slope must be very gradual. Collapse cannot occur in at least the first 10 billion years, and ensuring this amounts to picking the slope such that $m_{slope}^3 \la M_{Pl} H_0^2 \simeq 10^{-39}$ eV. If, thanks to technical naturalness, we saturate this bound, the universe is on the brink of collapse today. 

Once this is arranged, a remarkable consequence is that at some time prior to collapse the trigger field will dominate the evolution and inevitably lead to a stage of accelerated expansion of the universe. This gives a natural and immediate realization of one of the first ever models of quintessence \cite{andrei87} but without any direct fine tunings just to yield late acceleration alone. The scale of the accelerated expansion is controlled by the slope of the potential alone, because the global constraints (\ref{varsl}) constrain the initial value of the trigger field $\varphi_{in}$ to be moderately trans-Planckian.  Moreover, the prediction is robust in the sense that the initial condition is insensitive to radiative corrections. Thus the scale of accelerated expansion is automatically comparable to the scale controlling the age of the universe, and for our choice of $m_{slope}$ it is precisely {\it now}. This gives a new solution of the `Why Now?' problem. In short, if we see signs of cosmological collapse, we see acceleration. Conversely, the present epoch of acceleration may be evidence of impending doom. The duration of this accelerated expansion is not very long, lasting possibly only ${\cal O}(1)$ efolds. This is seen from the effective dark energy equation of state, which by (\ref{slowroll}) during this epoch is $w_{DE}+1\simeq 2 \epsilon \simeq 
{\cal O}(1) \frac{m_{slope}^6}{M_{Pl}^2 H_\dagger^4}  \simeq {\cal O}(1)\frac{H_0^4}{H_\dagger^4} $. Taking $H_\dagger$ to be  the Hubble scale at dark energy-matter equality,  
$w_{DE}$ can be measurably different from $-1$. 
Note also that since the universe is large, it should be spatially positively curved \cite{kp2}, with 
$\Omega_k < 0$. A detailed analysis to better quantify these predictions is certainly warranted.

\vskip.2cm

{\bf Acknowledgments}: 
We wish to thank to E. Copeland, G. D'Amico, A. Lawrence, A. Linde, A. Moss, J. Pearson, P. Saffin, D. Stefanyszyn, J. Terning, and A. G. Westphal and two `characters' in a pub in Woolsthorpe, UK 
for very useful discussions.
NK thanks the School of Physics and Astronomy, U. of Nottingham for hospitality in the course of this work.
NK is supported by the DOE Grant DE-SC0009999. AP was funded by a Royal Society URF.


\begin{thebibliography}{99}

\bibitem{kp1} 
N.~Kaloper and A.~Padilla,
Phys.\ Rev.\ Lett.\  {\bf 112}, 091304 (2014).
  
\bibitem{kp2} 
N.~Kaloper and A.~Padilla,
arXiv:1406.0711 [hep-th].
  
\bibitem{wein}
S.~Weinberg,
Rev.\ Mod.\ Phys.\  {\bf 61}, 1 (1989).

\bibitem{zeldovich} 
Y.~B.~Zeldovich,
JETP Lett.\  {\bf 6}, 316 (1967)
[Pisma Zh.\ Eksp.\ Teor.\ Fiz.\  {\bf 6}, 883 (1967)];
Sov.\ Phys.\ Usp.\  {\bf 11}, 381 (1968).

\bibitem{wilczek} 
F.~Wilczek,
Phys.\ Rept.\  {\bf 104}, 143 (1984).

\bibitem{lindelinder}
R.~Kallosh, J.~Kratochvil, A.~D.~Linde, E.~V.~Linder and M.~Shmakova,
JCAP {\bf 0310}, 015 (2003).
  
\bibitem{giudice} 
G.~Degrassi, S.~Di Vita, J.~Elias-Miro, J.~R.~Espinosa, G.~F.~Giudice, G.~Isidori and A.~Strumia,
JHEP {\bf 1208}, 098 (2012).

\bibitem{sala}
  D.~Buttazzo, G.~Degrassi, P.~P.~Giardino, G.~F.~Giudice, F.~Sala, A.~Salvio and A.~Strumia,
  JHEP {\bf 1312} (2013) 089
  [arXiv:1307.3536 [hep-ph]].
  
\bibitem{whynow}
 P.~J.~Steinhardt, in 
 ``Critical Problems in Physics'', edited by V.~L.~Fitch, and D.~R.~Marlow,
Princeton University Press, Princeton, NJ,
 1997.
 
 
\bibitem{Avelinonow}
  P.~P.~Avelino,
  Phys.\ Lett.\ B {\bf 611} (2005) 15
  [astro-ph/0411033].
  A.~Barreira and P.~P.~Avelino,
  Phys.\ Rev.\ D {\bf 83} (2011) 103001
  [arXiv:1103.2401 [astro-ph.CO]].


\bibitem{waldnohair}
R.~M.~Wald,
Phys.\ Rev.\ D {\bf 28}, 2118 (1983).
  
\bibitem{selft} 
N.~Arkani-Hamed, S.~Dimopoulos, N.~Kaloper and R.~Sundrum,
Phys.\ Lett.\ B {\bf 480}, 193 (2000);
S.~Kachru, M.~B.~Schulz and E.~Silverstein,
Phys.\ Rev.\ D {\bf 62}, 045021 (2000).

\bibitem{Avelinoseq}
  P.~P.~Avelino,
  Phys.\ Rev.\ D {\bf 90} (2014) 10,  103523
  [arXiv:1410.4555 [gr-qc]].


\bibitem{andrei87}
 A. Linde, ``Inflation and quantum cosmology", p. 604, in ``300 Years of Gravitation'',
 ed. S. Hawking and W. Israel, Cambridge Univ. Press, 1987.
 
 
 
 \end{thebibliography}
\end{document}